\documentclass{article}

\PassOptionsToPackage{numbers, compress}{natbib}
\usepackage[final]{nips_2017}

\usepackage[utf8]{inputenc} 
\usepackage[T1]{fontenc}    
\usepackage{hyperref}       
\usepackage{url}            
\usepackage{booktabs}       
\usepackage{amsfonts}       
\usepackage{nicefrac}       
\usepackage{microtype}      
\usepackage{ntheorem}       
\usepackage{cleveref}       
\usepackage{graphicx}       
\usepackage{subcaption}        
\usepackage[table,xcdraw]{xcolor} 
\usepackage{tikz}
\usetikzlibrary{shapes,backgrounds}
\usepackage{float}

\graphicspath{{./figures/}}

\title{Visual Display and Retrieval of Music Information}

\author{
  Rafael Valle \\
  CNMAT, UC Berkeley \\
  \texttt{rafaelvalle@berkeley.edu} \\
}

\begin{document}
\maketitle
\begin{abstract}
    This paper describes computational methods for the visual display and
    analysis of music information. We provide a concise description of software,
    music descriptors and data visualization techniques commonly used in music
    information retrieval. Finally, we provide use cases where the described
    software, descriptors and visualizations are showcased.
\end{abstract}
\section{Introduction} \label{sec:introduction}
In the 21st century, the most common methodologies for music analysis are the
visual study of the musical score and the aural and perceptual investigation of
recordings\cite{adorno1982}. In analysis courses, analysts are usually
interested in creating a holistic description of a piece of music by analysing
relationships and patterns, local and global, among descriptors notated on the
score or perceived through listening.

Another less common but extremelly powerful methodology for music analysis is
computer aided music analysis~\cite{downie2003}, also known as computational
musicology or music information retrieval (MIR). This methodology is very
important for companies like Spotify, Pandora\cite{prockup2015}\footnote{Pandora
combines human and computer aided music analysis.} and Gracenote~\cite{liem2011}
who rely on algorithms and machine learning to extract information from music
and listener behavior. Among musicians, computer aided music analysis and
composition has been increasing, specially amid composers; among musicologists,
Computational musicology is less common, perhaps due to the lack of related
courses and the costs of learning a programming language and data analysis.

Information retrieval with computational means has played an important role in
helping us learn about and advance our understanding of music and many other
things~\cite{singhal2001}. For example, it empowers the analysis of hundreds of
thousands of musical compositions with the speed of a mouse click and at the
cost of few GPU cycles! It enables types of interesting analysis whose execution
would be cumbersome for humans, such as understanding the distribution of events
in a piece, comparing multiple performances of the same piece and understanding
relationships between timbres, to cite a few. Similarly to other fields, it also
empowers analysis without a precedent in history that can potentially lead to
new discoveries! Last and equally important, computational musicology allows for
the quantification of the musical analysis, allowing for more systematic and
less subjective conclusions\footnote{with the caveat that the analysis method is
also subjective}.

Interestingly, a recent study\cite{cartwright2017} provides evidence of the
importance of data visualization in music and shows that, in the context of
crowdsourcing music annotations, "\textit{more complex audio scenes result in
lower annotator agreement, and spectrogram visualizations are superior in
producing higher quality annotations at lower cost of time and human labor.}"

\section{Information Retrieval}\label{sec:retrieval}
This section gives a brief description of music information retrieval in the
form of pitch and timbre descriptors. It focuses on descriptors that are not
easily, if possible at all, \textbf{manually} extracted from musical scores, audio
recordings and symbolic music, e.g MIDI files or OSC. We forward the reader to
\cite{typke2005,peeters2004} for a survey of MIR systems and commonly used audio
descriptions. 

\subsection{Timbre}
\begin{figure*}[!h]
    \caption{Beat-aligned Power Spectrogram extracted from a recording of Gerard
    Grisey's Partiels. Vertical axis maps to frequency bin in hz, horizontal
    axis maps to time.}
    \centering
    \includegraphics[width=\textwidth]{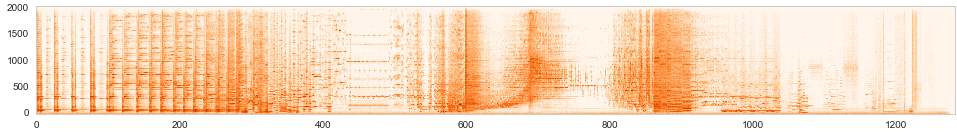}
    \label{fig:STFT}
\end{figure*}

Computing power can be used to project an audio signal from a time domain
representation, e. g. amplitude over time, to a frequency domain representation,
e. g. magnitude of frequency bin over time. There are many such projections and
in the field of music information retrieval commonly used projections for timbre
are the Spectrogram, computed with the Short Time Fourier Transform
(STFT)\cite{allen1977}, and the Mel-Frequency Cepstral Coefficients
(MFCC)\cite{logan2000}. When computing such frequency domain representations,
the time domain signal is usually divided into overlapping analysis windows. For
each analysis window, the Spectrogram, as shown in Figure~\ref{fig:STFT},
provides the magnitude of the frequency bins.  

\begin{figure*}[!h]
    \caption{Beat-aligned MFCC extracted from a recording of Gerard Grisey's
    Partiels. Vertical axis maps to the mel-frequency cepstral coefficient,
    horizontal axis maps to time.}
    \centering
    \includegraphics[width=\textwidth]{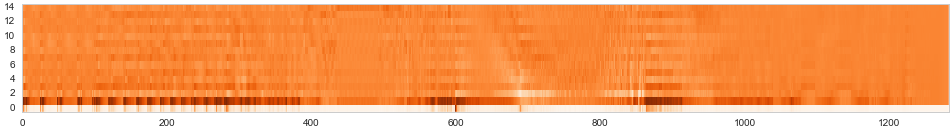}
    \label{fig:MFCC}
\end{figure*}

Mel-Frequency Cepstral Coefficients (MFCC), shown in Figure~\ref{fig:MFCC} are
probably the the most used descriptors in Automatic Speech Recognition
(ASR)~\cite{hinton2012}. The MFCC arguably mimics some parts of human speech
production and speech perception, specially the logarithmic perception of
loudness and pitch in the human auditory system. 

\begin{figure*}[!h]
    \caption{Beat-Aligned CQT extracted from a recording of Grisey's
    Partiel. Vertical axis maps to notes, horizontal axis maps to time.}
    \centering
    \includegraphics[width=\textwidth]{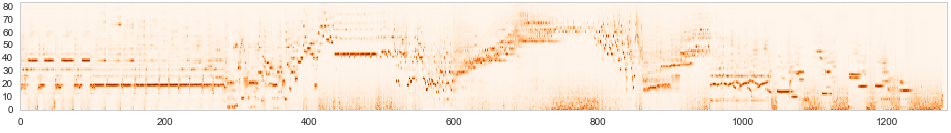}
    \label{fig:CQT}
\end{figure*}

\begin{figure*}[!h]
    \caption{Beat-aligned Chromagram extracted from a recording of Grisey's
    Partiel. Vertical axis maps to pitch classes, 0 is C, horizontal axis maps to time.}
    \centering
    \includegraphics[width=\textwidth]{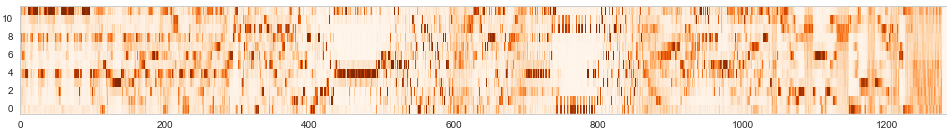}
    \label{fig:Chroma}
\end{figure*}

\subsection{Pitch}
There are two pitch based features that are commonly used in MIR, being the
Constant Q Transform (CQT) and the Chroma. The CQT and the Spectrogram are
closely related, in which they are both a bank of filters that measure the
amount of energy on each frequency bin over time. In contrast to the
Spectrogram, the CQT is usually interpreted as a Piano Roll given its
geometrically spaced frequencies that, informally speaking, resemble the
equal-tempered system and can be mapped to specific notes, e. g. the notes on a
piano. 

The Chroma is also referred to as pitch class profile. It is closely related to
the CQT~\cite{bello2005} in which it measures the energy of specific pitches but
disregards their octave, that is, Chroma measures the amount of energy on each
pitch class\footnote{Although these pitch classes normally follow equal
temperament, other temperaments are possible.}. Chromagrams such as the one in
Figure \ref{fig:Chroma}, display a sequence Chroma, also known as Chromagram.
\section{Information Visualization}\label{sec:visualization}
In this section we provide an overview of visualization techniques that are
commonly used in data analysis and that can facilitate the understanding of musical data.
We forward the interested reader to Edward Tufte's book~\cite{tufte1983}.

\subsection{Sets}
    In mathematics, the term set refers to a finite or infinite collection of
    objects in which order is not relevant and existance of an item in the set
    is more important than its exact count. Sets are visualized using Venn
    diagrams that describe the composition of each set and possible
    relationships. Basic set operations include union, intersection, complement
    and difference. Musical set theory has been considerably used in music
    composition~\cite{castine1994}, including computer-aided composition, and
    pitch analysis. The following diagram shows the pitch classes of the C
    major, orange, and C minor, blue, scale.
    \begin{center}
    \begin{tikzpicture}
        \tikzset{venn circle/.style={draw,circle,minimum width=2.7cm,fill=#1,opacity=0.5}}
        \node [venn circle = orange,align=center] (A) at (0,0) {\tiny 4\\
                                                                \tiny 9\\
                                                                \tiny 11};
        \node [venn circle = blue,align=center] (B) at (0:1.7cm) {\tiny 3\\
                                                                  \tiny 8\\
                                                                  \tiny 10};
        \node[align=center] at (barycentric cs:A=1/2,B=1/2)  {\tiny 0\\
                                                              \tiny 2\\
                                                              \tiny 5\\
                                                              \tiny 7}; 
        \end{tikzpicture}  
    \end{center}

\subsection{Distributions and Histograms}
    Similar to sets, distributions refers to a finite or infinite collection of
    objects in which order is not relevant. Unlike sets, distributions display
    the quantity of each item or numerical range, which can be represented as
    absolute quantity or scaled to represent the frequency of each item or
    numerical range. Bar graphs, pie charts, density plots, box plots are data
    visualizations commonly used by scientists to analyze the occurence of
    discrete and continuous objects.

    \begin{figure}[!ht]
        \caption{Pie Chart describing the usage in movies of Naxos recording of
        Mozart's music.} 
        \centering
        \includegraphics[width=0.28\textwidth]{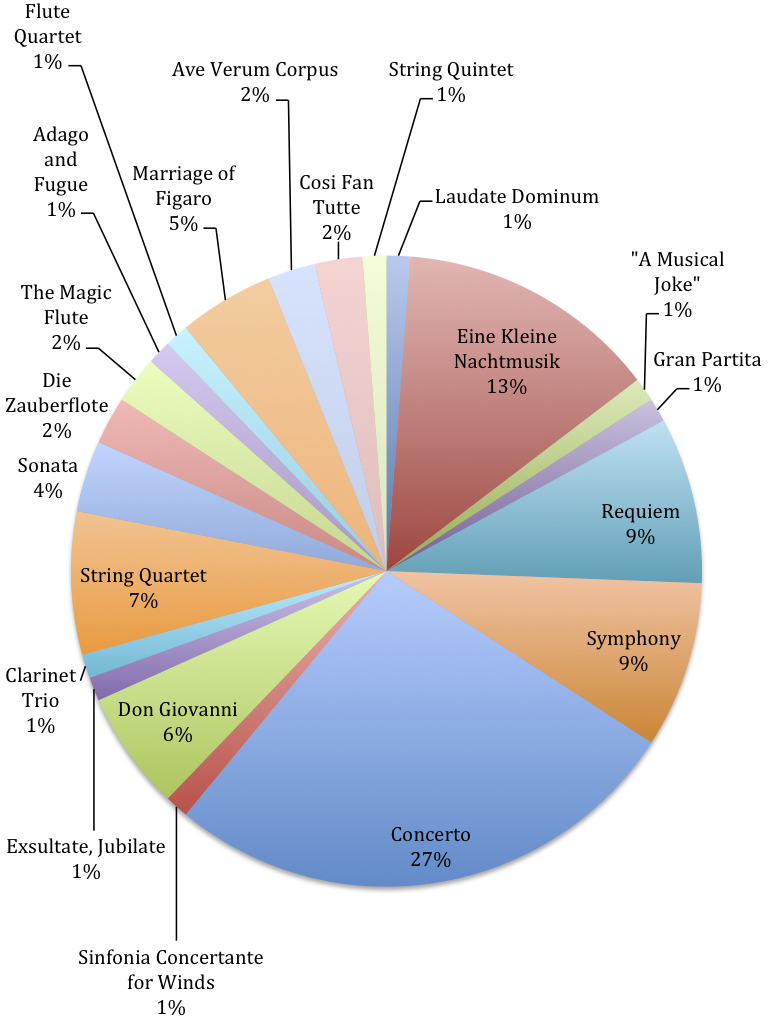}
        \label{fig:Mozart-Movie-Pie}
    \end{figure}

    The pie chart is a useful static visualization of proportions between data
    items where the area of each item is dependent on the area of other items
    and the total area must sum to the total area of the shape in which the data
    is displayed. In Figure~\ref{fig:Mozart-Movie-Pie} we provide one example
    found in a music blog that was the outcome of a class
    project~\cite{handmusicblog}.

\begin{figure}[!ht]
        \caption{Histogram of Pitch Classes extracted from The Beatle's Birthday
        song. Note how the most frequent notes relate to D mixolidian.}
        \centering
        \includegraphics[width=0.5\textwidth]{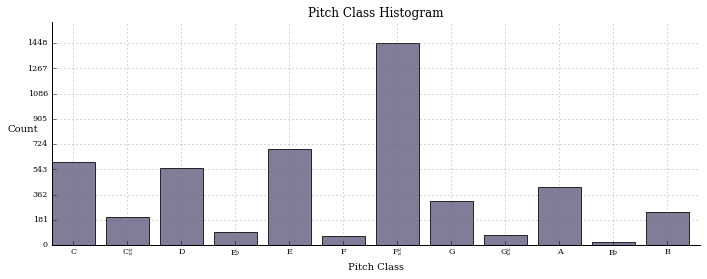}
        \label{fig:PCP}
    \end{figure}

    For dynamic visualization, pie charts become cumbersome because the location
    and area of each item is dependent on the other items. A better tool for
    such cases is the bar graph, where the location of each item is fixed and
    their proportions is perceived by comparing the heights of each bar. Figure
    \ref{fig:PCP} shows an example of a pitch class histogram, represented as a
    bar graph, extracted using the software Music 21\cite{cuthbert2010} from a
    MIDI rendering of Birthday by The Beatles.

\begin{figure}[!ht]
        \caption{Boxplot of recorded song duration given decade. Note the increase in
        song duration given decade.}
        \centering
        \includegraphics[width=0.5\textwidth]{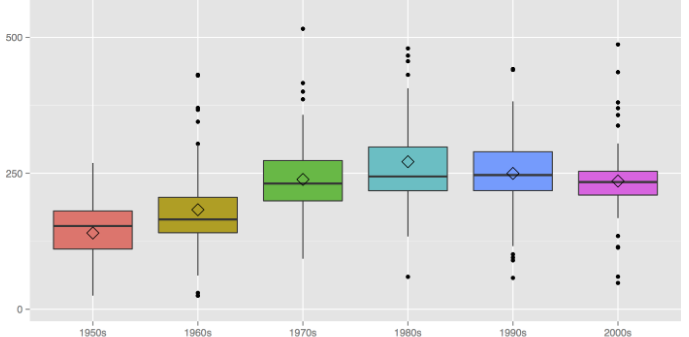}
        \label{fig:Hits-Duration-Pie}
    \end{figure}

    Boxplots are useful to visualize the distribution of quantities, specially
    quartiles, and to visualize quantities that deviate from the average
    quantity by some value. Box plots provide information about the spread and
    skewness of the data being analyzed. Multiple box plots side-by-side can
    facilitate the comparisson of multiple distributions, as shown in
    Figure~\ref{fig:Hits-Duration-Pie} from~\cite{naresh2016}.

\subsection{Quantitative relationships}
    Quantitative relationships between descriptors can be visualized with
    graphs, normally limited to 3 dimensions, as it is not easy for humans to
    visualize data in more than just a few dimensions.  Quantitative
    relationships can clarify the understanding of the correlation, linear or
    non-linear, between two or more variables. In music, it is very common for
    one of these variables to be time.

    The numerical relationships between descriptors can be summarized with
    single quantities or equations, e. g. correlation, the slope of a line or
    the average length of events. This information can be visualized with
    matrices and graphs, line or scatter. Matrices and graphs can provide fast
    access to understanding information and relationships between descriptors.
    Consider visualizing the rhtyhmic density of a solo over time, or a matrix
    that displays the correlation between music styles: whereas the first
    provides a visualization that can describe form according to rhythmic
    descriptors, the second  can provide visualization of hidden style
    groupings. 
    
\begin{figure}[ht]
        \caption{Matrix and polar visualization of correlation between variables
        0, 1 and 2. Although the correlation matrix seems correct, a polar
        visualization immediately shows that the correlations are not valid.}
        \centering
        \includegraphics[width=0.26\textwidth]{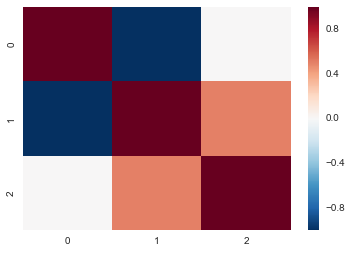}
        \includegraphics[width=0.21\textwidth]{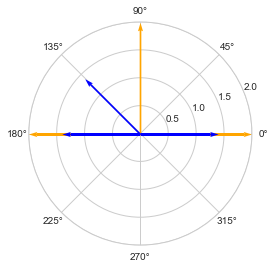}
        \label{fig:Correlation}
    \end{figure}

    Quantitative information can also be described using Polar notation. Polar
    notation is commonly used to display operations between vectors, to display
    phase in audio signals and, naturally, time in watches. In
    figure~\ref{fig:Correlation} we illustrate the benefit of visualizing
    correlations with polar notation to validate correlation matrices.

\subsection{Qualitative relationships}
    Qualitative relationships between descriptors can be visualized with
    diagrams, for example. Borrowing from mathematics, one can describe
    qualitative relationships with a graph that is comprised by a set of nodes
    and connections between nodes called edges. Each node can describe a state,
    e. g. a quantity, object or action and the edges between the states can
    describe a qualitative or quantitative relationship between the states, e.g.
    a direct transition or ownership. 
    
\begin{figure}[ht]
        \caption{1945.}
        \centering
        \includegraphics[width=0.5\textwidth]{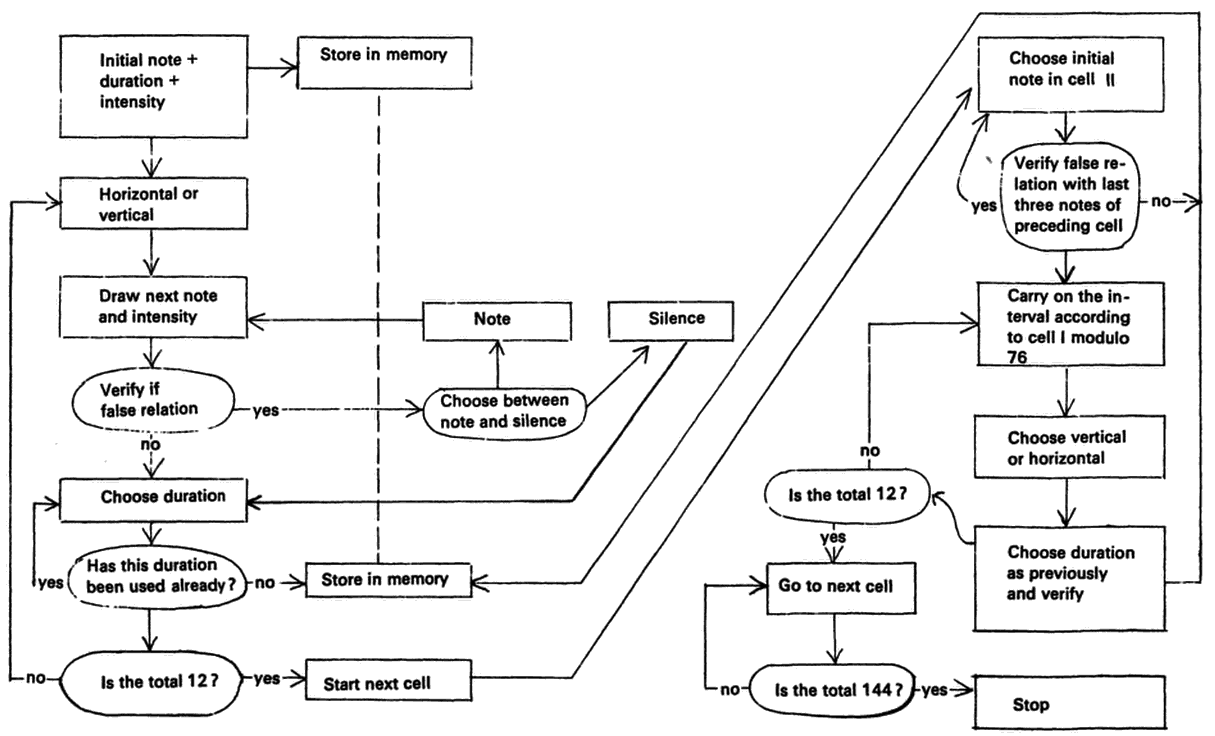}
        \label{fig:Philippot-Double-Graph}
    \end{figure}

    In music analysis, it is common to represent the states as pitches,
    durations or some symbollic encoding of audio; the edges usually represent
    possible transition from one state to another. Note that a transition does
    not need to be bidirectional, that is, a transition from A to B does not
    imply a transition from B to A. Figure~\ref{fig:Philippot-Double-Graph}
    shows a graph used by Michel Philippot to compose the first movement of his
    piece \textit{Composition} for Double Orchestra.

\section{Software and Datasets}\label{sec:software}
In this section we introduce free computer programs and libraries that can be
used in computer aided music analysis. Although most of these programs are
dedicated to either symbolic music formats, e.g. MIDI, or Audio, there a few
exceptions that operate on both domains.

In the past decades, several programs for computer aided musVic analysis and
composition have been designed. IRCAM's \textit{Open Music} and its ancestor
\textit{Music V} are such programs that were used in computer aided composition
and analysis~\cite{assayag1999}, though their focus has been on music
composition. David Huron's \textit{Humdrum}~\cite{huron2002} is still active
nowadays and is one of the pioneer's in musical analysis using algorithms based
on human cognition. \textit{Music 21}~\cite{cuthbert2010} by Michael Cuthbert
from MIT is a python based software for symbolic music analysis with a rather
active user base. In addition to supporting multiple file formats, such as XML,
humdrum, and MIDI, it has a vast amount of routines that produce low level
descriptors and analytical results, related to scales, chords, key and harmonic
analysis, pitch and duration summaries and others. Similar to \textit{Music 21},
\textit{pretty\_midi}~\cite{raffel2014} by Google Brain's Colin Raffel is a
light-weight and very efficient python library for extracting low level
descriptors such as key, meter, pitch and duration from MIDI files or computing
descriptors that are usually associated with audio, such as CQT and Chromagrams.

\textit{Librosa}~\cite{mcfee2015}, \textit{madmom}~\cite{bock2016},
\textit{essentia}~\cite{bogdanov2013} and \textit{marsyas}~\cite{tzanetakis2000}
are libraries for retrieving and visualizing information from audio. They are
associated with the respective universities or research labs they come
from\footnote{Labrosa, OFAI, MTG}. These libraries provide a large collection of
routines that allow users to extract low level descriptors, analyze audio and
build statistical models.

Although the programs mentioned in this section require users to have some
knowledge of computer programming, this initial cost is usually amortized by
end-to-end tutorials, where the users learn, for example, how to write code to
extract the key from an audio file or to create a toy model for genre
classification using audio data.

\section{Applications}\label{sec:applications}
This section showcases a few techniques used in music information retrieval and
visualization. We forward the interested readers to the proceedings of the
International Society for Music Information Retrieval (ISMIR), conferences and
the ISMIR mailing list.

\subsection{Data abstraction}
\begin{figure}[ht!]
        \caption{Timbre density in Xenakis' ST.}
       \centering
        \includegraphics[width=0.5\textwidth]{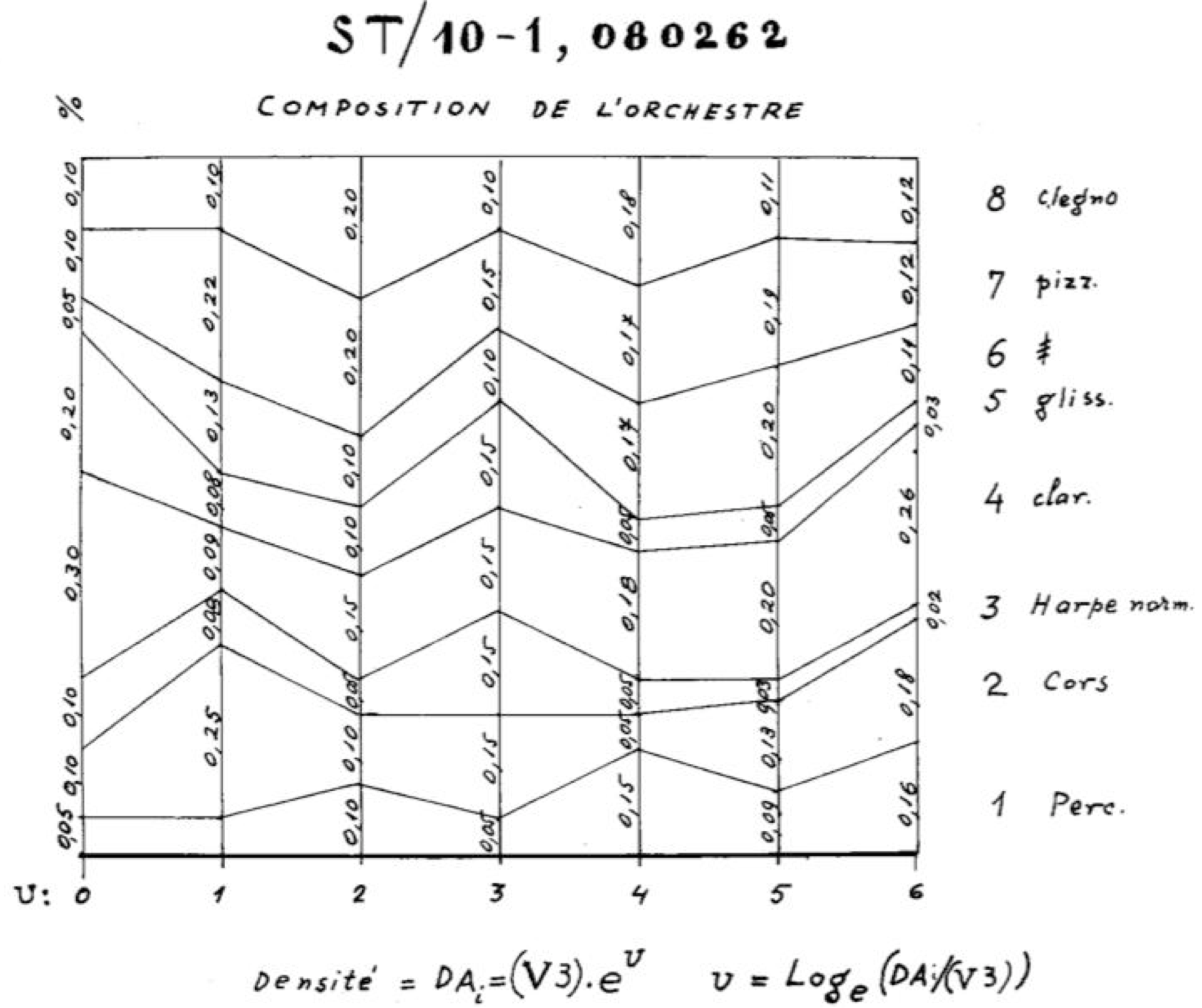}
        \label{fig:Xenakis-ST-density}
    \end{figure}

    We informally describe a data abstraction as a function, e.g. an equation or
    algorithm, that maps an abstract source object onto another object that
    summarizes the information on the source object, e. g.  timbre density is a
    data abstraction of timbre as shown in Figure~\ref{fig:Xenakis-ST-density}.

    A common data abstraction in computational music information retrieval are
    pitch profiles, which refer to histograms or distributions of pitches. For
    example, Katy C. Noland and Mark B. Sandler have computationally summarized
    the energy on pitch classes in major or minor pieces of Bach's Well Tempered
    Clavier and used each summary to create pitch class profiles that were
    associated with the major and minor modes\cite{noland2007}. Following
    Noland's approach, other researchers have used MIDI files to compute the
    frequency of transitions between pitch classes and create pitch class
    transition profiles that were associated with the key and mode from which
    they were extracted. In addition to providing a visualization of the
    distribution of pitch classes, such as Figure~\ref{fig:PCP}, and pitch
    transition, these distribution have been used as features to build key,
    harmony and chord classifiers.

\subsection{Specifications and Automata}
\begin{figure}[!ht]
    \centering
    \includegraphics[width=\columnwidth]{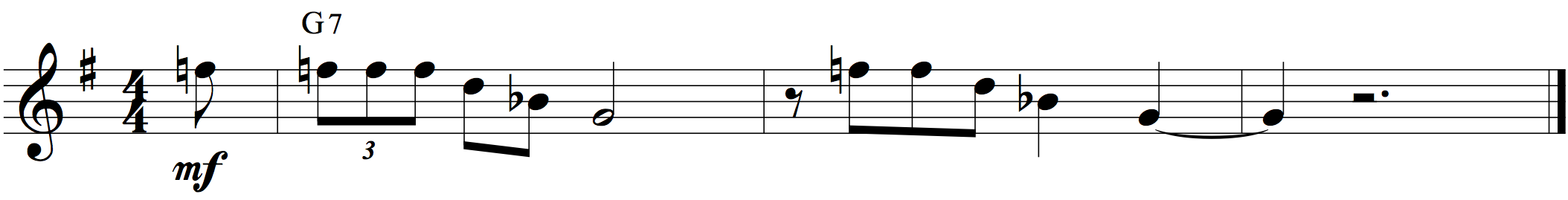}
    \caption{First phrase of Crossroads Blues by Robert Johnson as transcribed in the Real Book of Blues.  The transition from chord degree 10 (note f) to chord degree 7 (note d) is always
        preceded by two or several occurrences of chord degree 10.}
    \label{fig:crossroads_blues}
\end{figure}

\begin{figure}[ht!]
    \centering
    \includegraphics[width=.85\linewidth]{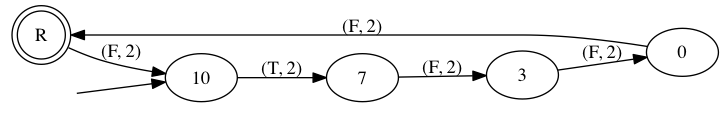}
    \caption{Pattern graph learned on the chord degree feature (interval from root) extracted from the phrase in Fig.~\ref{fig:crossroads_blues}. The \textbf{F} pattern between chord degrees 10 and 7 has been merged into the pattern 10 \textbf{T} 7.}
    \label{fig:graph_example}
\end{figure}

    Figure~\ref{fig:Philippot-Double-Graph} describes a graph that has been used
    to compose a piece of music. The graph in Figure~\ref{fig:graph_example}
    can be used to describe the specifications of a song, loosely speaking, the
    patterns that are valid or characteristic of a song. More formally, a
    pattern graph is a labelled directed multigraph whose nodes are values of
    the descriptors it refers to. A node can be labelled as a starting node, an
    ending node, or neither. Edges can be labelled with a word that describes
    the pattern between nodes and a count indicating how many times the pattern
    occurred. For example, an edge (a, b) labelled (F, 2) in the pattern graph
    means the pattern a F b occurred 2 times. The complete example of a pattern
    graph in Figure \ref{fig:graph_example}, where we have indicated starting
    nodes with an unlabelled incoming arrow and ending nodes with a double
    circle, illustrates specifications a blues song in
    Figure~\ref{fig:crossroads_blues}.


\begin{figure}[!ht]
    \centering
    \includegraphics[width=.4\linewidth]{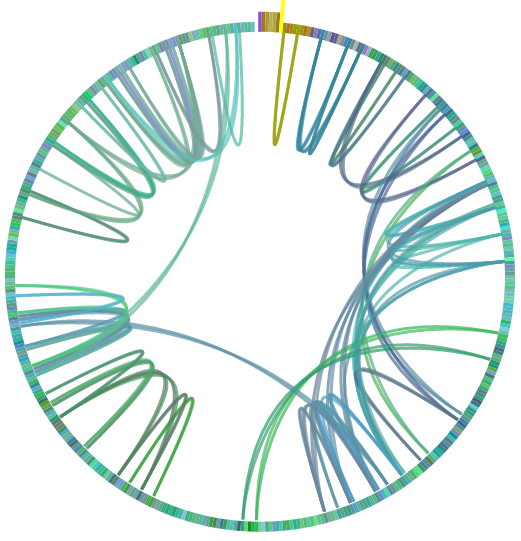}
    \caption{Infinite Jukebox built on the song Supersticion by Stevie Wonder}
    \label{fig:InfiniteJukebox}
\end{figure}

Graphs visualizations are common amongst researchers working with graphical
models and oracles. The Infinite Jukebox shown in
Figure~\ref{fig:InfiniteJukebox} is an example of such. Note that as the number
of states and edges increases, such graphs become significantly complex,
requiring careful attention during visual analysis or making it unfeasible.

\subsection{Structure Analysis}
\subsubsection{3D Plots}
    A visualization of the CQT over time, such as Figure~\ref{fig:CQT}, can
    facilitate the visualization of the structure of a piece. Another
    visualization that provides information about the structure of a piece are
    3D plots, which require mechanisms\footnote{Principal Component
    Analysis(PCA), Singular Value Decomposition (SVD), etc...} to project the
    data abstraction onto a 3D space and to embed temporal information into the
    plot. 

\begin{figure}[!ht]
        \caption{Grisey's Partiels SVD(CQT)}
        \centering
        \includegraphics[width=0.2\textwidth]{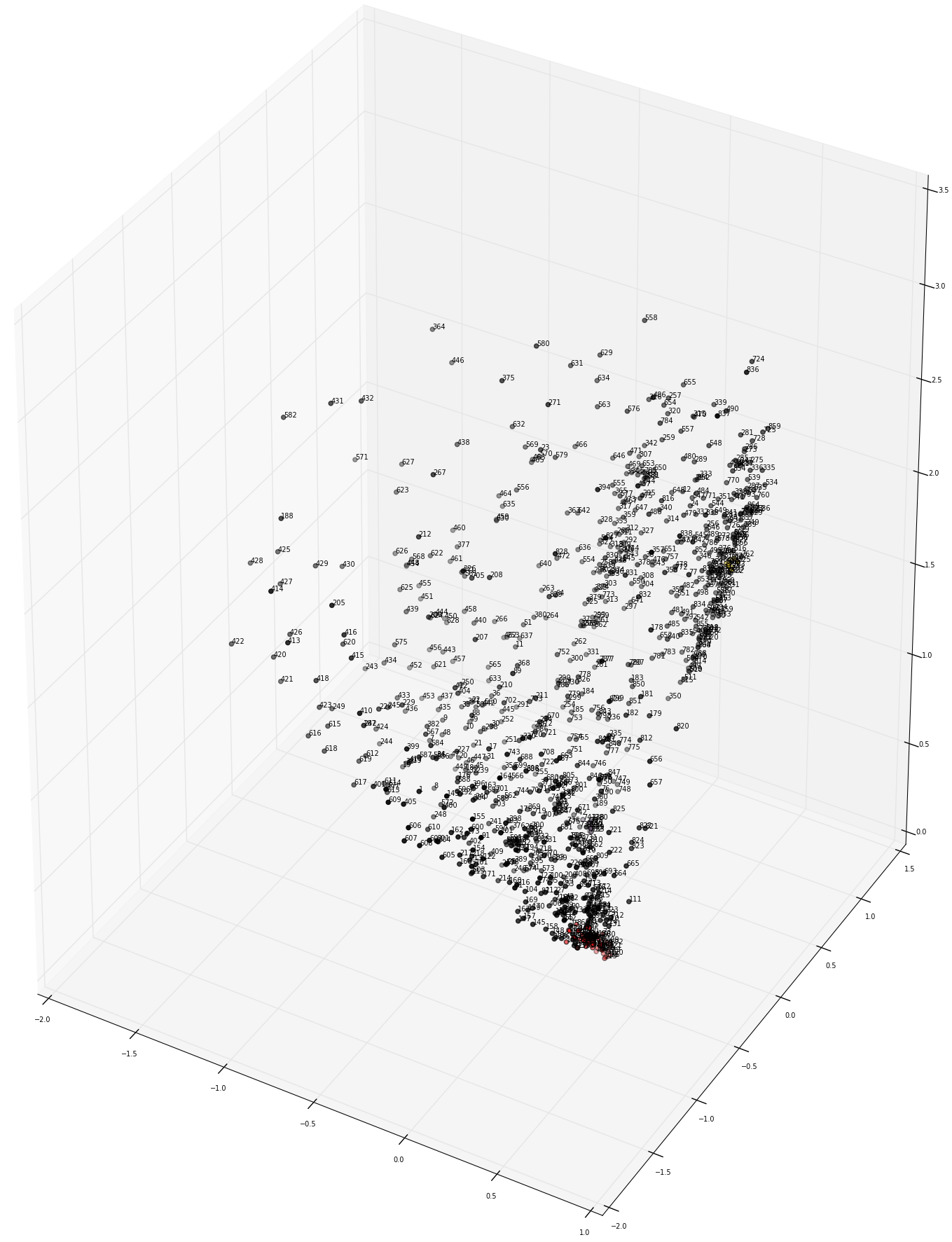}
        \label{fig:Grisey-SVD-CQT}
    \end{figure}

    An explicit mechanism to embed temporal information is to annotate each data
    point in the plot with its order of appeareance. With this, the order can be
    retrieved by simply following the numerical order.
    Figure~\ref{fig:Grisey-SVD-CQT} shows such plot and the inevitable
    cluttering with many data points.
    
\begin{figure}[!ht]
        \caption{Grisey's Partiels LPF(SVD(CQT))}
        \centering
        \includegraphics[width=0.2\textwidth]{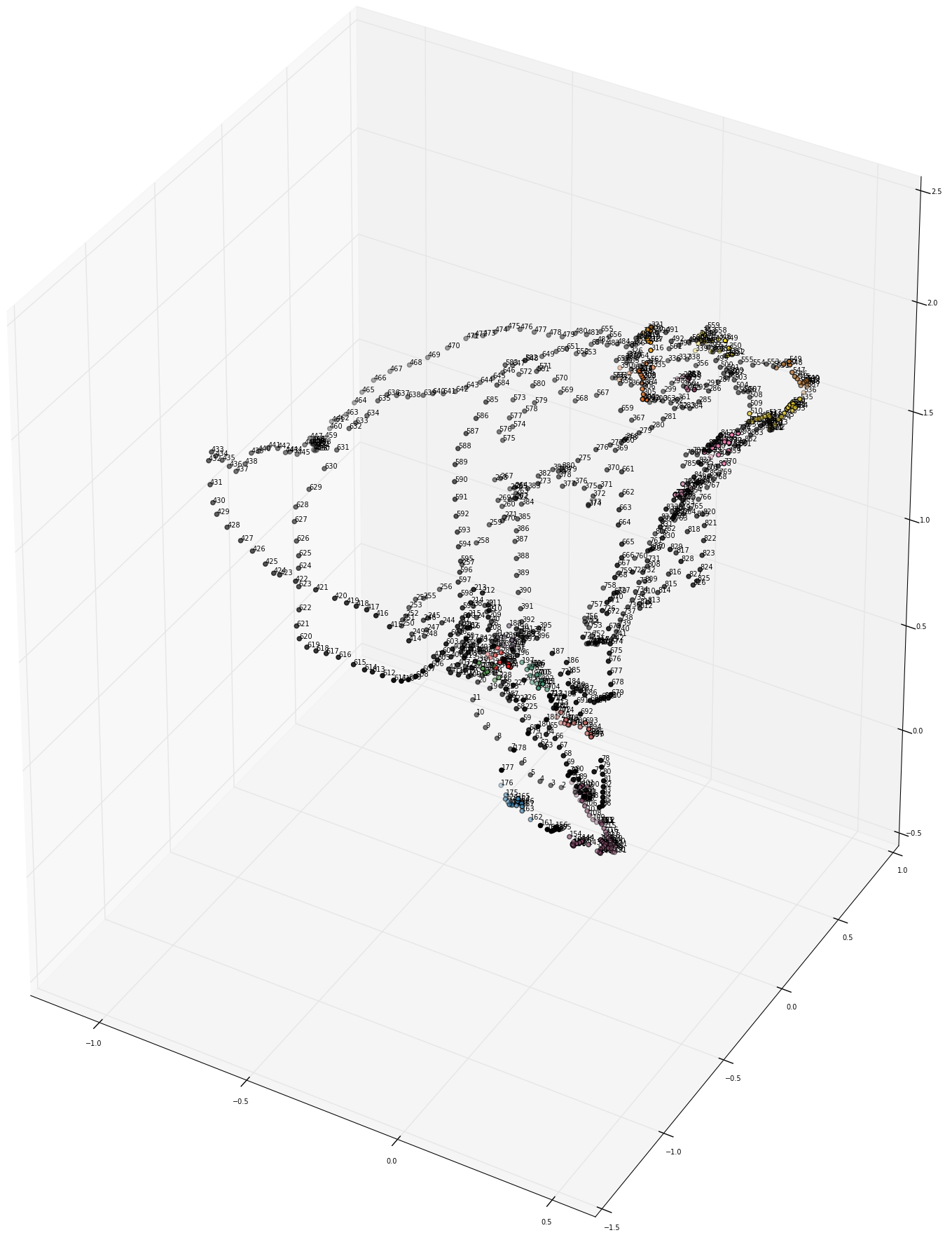}
        \label{fig:Grisey-LPF-SVD-CQT}
    \end{figure}

    An implicit mechanism to embed temporal information consists in low-pass
    filtering (LPF) the data: LPFs can be used to \textit{bleed} information
    from previous data frames to the current data frame. In a plot visualization
    such as Figure~\ref{fig:Grisey-LPF-SVD-CQT}, this creates a trajectory that
    produces an implicit visualization of the temporal sequence of the data
    points.
    
\subsubsection{Color sequence}
    Features like the CQT and Chroma can be used by the music specialist to
    identify patterns in the music. For the enthusiast or novice, a color
    sequence might be more indicative of similarities and form in music. 
    
    Dimensionality reduction can be used to project any feature, for example the
    CQT with 88 dimensions, to the number of dimensions of a color scheme, for
    example the RGB color scheme has 3 dimensions: red, green and blue.

\begin{figure*}[ht!]
        \caption{Grisey's Partiels Color Sequence Based on LPF(SVD(CQT)).}
        \centering
        \includegraphics[width=\textwidth]{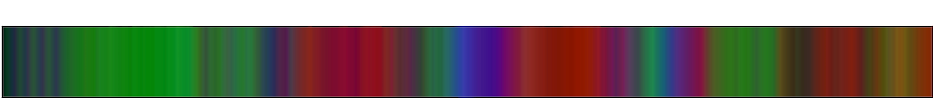}
        \label{fig:color_sequence}
    \end{figure*}

    In Figure~\ref{fig:color_sequence}, we provide a visualization where CQT
    features where projected onto a 3D space. After proper scalling, the
    projected data can then be interpreted and visualized as an image in RGB,
    where the temporal sequence of colors is directly mapped to the temporal
    sequence of music, and the similarity in color corresponds to similarity in
    music, as described by the feature projected onto the color space.

\subsubsection{Self-similarity matrices}
    Self-Similarity Matrices have been used as the main feature for
    computational structure segmentation. They describe the similarity, as
    described by some distance function, between all frames~\footnote{Other
    units can be used.} in a music piece. The self-similarity matrix is square
    and symmetric on the diagonals, that is, the matrix can be divided into two
    equal triangles.

    \begin{figure}[ht!]
        \caption{Grisey's Partiel Self Similarity Matrix computed using
        Beat-Aligned CQT}
        \centering
        \includegraphics[width=0.5\textwidth]{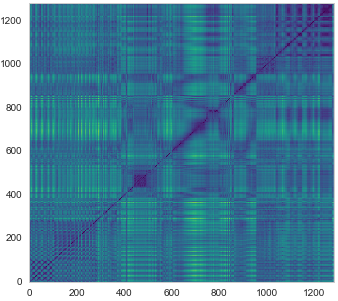}
        \label{fog:SSM}
    \end{figure}    

\subsection{Similarity}
    Similarity between musical entities is an important subject in music
    research. Computation enables systematic quantification of similarity
    between musical entities as described by the features and distance measures
    designed.

    \subsubsection{Feature choice and distance measure}
        To illustrate the importance of the feature choice, let's consider a
        musical melody of reference, its octave above and diminished fifth above
        transposition, named $C_r$, $C_o$ and $C_f$ respectively. Whereas some
        people people would claim that $C_o$ is the most similar to $C_r$
        because the octave is a more consonant interval, other people would say
        that $C_f$ is the most similar because it has a smaller absolute
        distance. If, however, the feature used was the sequence of intervals,
        the melodies would be considered similar to each other.

        Now, to illustrate the importance of the distance measure, let's
        consider tree pitch profiles $P_r$ (C, E, G at 50\% volume), $P_1$ (no
        notes), and $P_2$(C, E, G at 100\% volume). Let's define our distance
        measure as the Euclidean distance, $d(x, y) = \sqrt{\sum_{i=1}^n(x_i -
        y_i)^2}$. Since the euclidean distance measures the point to point
        distance between the frequency of each pitch class in the pitch class
        profiles, the distances d($P_1$, $P_r$) and d($P_2$, $P_r$) are exactly
        the same. On the other hand, musicians would use a distance measure
        that is more similar to computing the correlation between profiles and
        say that $P_2$ is closer to $P_r$ than $P_1$)

    \subsubsection{Similarity Spaces}
        Given features computed over time windows, i.e. frames, that describe
        musical entities, computational means can be used to visualize this data
        and quantify the distances between these frames. The frames could
        provide, for example, per beat information about some spectral
        descriptor, e.g. brightness, of songs. One could then plot these frames
        and their respective brightness and compare, for example, different
        songs or sections of the same song.
        
        As mentioned before, visualization of data in more than 3 dimensions is hard
        and will require a mechanism, dimensionality reduction or creative plotting,
        to reduce the dimensionality of the data or find other ways of plotting it,
        e.g. using color and shapes to represent other dimensions in the data.

        There are several methods for performing dimensionality reduction, including
        Multidimensional Scaling (MDS), Principal Component Analysis (PCA) and t-SNE, to cite a
        few. In the late 70s, David Wessel used Multidimesional Scaling to create a
        timbre space where the distance between points were related to the timbral
        distance between instruments\cite{wessel1979}. 
        
        \begin{figure}[ht!]
            \centering
            \includegraphics[width=.5\linewidth]{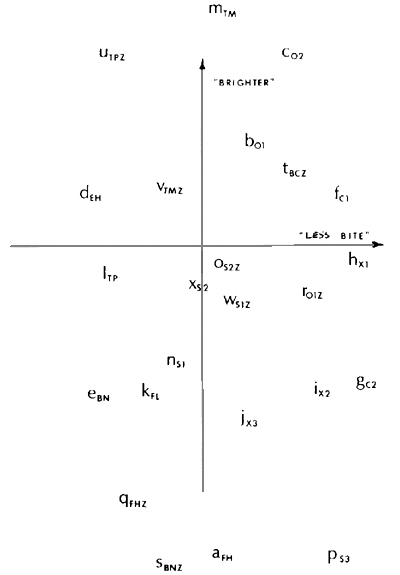}
            \caption{Two-dimensional timbre space representation of 24 instrument-like sounds obtained from Grey.}
            \label{fig:Wessel-TimbreSpace}
        \end{figure}

        Unlike MDS and PCA, t-SNE is a dimensionality reduction technique that,
        informally speaking, tries to preserve in the projected low-dimensional
        space the distances between points in high-dimensional space.

\begin{figure}[!hb]
            \centering
            \includegraphics[width=\linewidth]{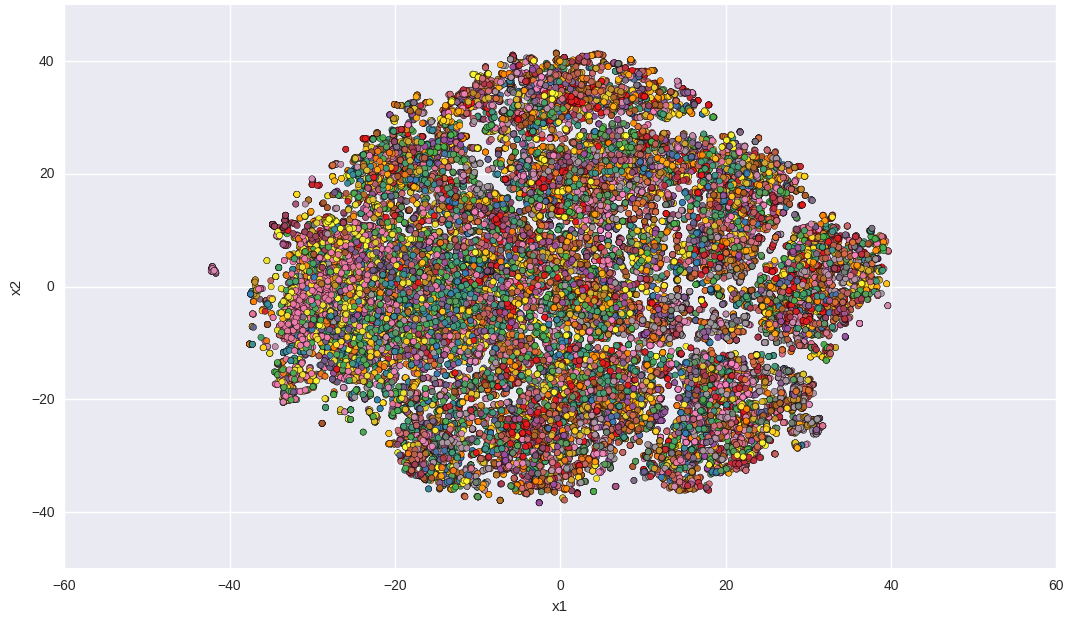}
            \caption{Frame Level Projection}
            \label{fig:FrameLevel}
        \end{figure}

\begin{figure}[!hb]
            \centering
            \includegraphics[width=\linewidth]{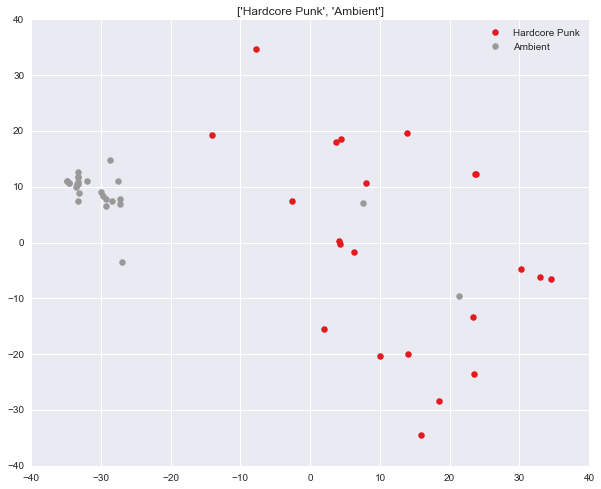}
            \caption{Hardcore and Ambient Projection}
            \label{fig:HCAmbient}
        \end{figure}

        Figure~\ref{fig:FrameLevel} shows a 2D t-SNE projection of feature
        frames from thousands of songs. Figure~\ref{fig:HCAmbient} uses the same
        data as Figure~\ref{fig:FrameLevel} and provides a visualization of
        median aggregated frames of Hardcore Punk and Ambient songs. Note the
        clear separation between styles and that the distance between points is
        a measure of similarity between songs. 

\section{Conclusions}\label{sec:conclusions}
This paper summarized a few computational strategies for music information
retrieval and visualization, including commonly use features and software
libraries.. It provided concrete examples of how visualization can be used to
retrieve information from music, including abstractions, specifications,
structure and similarity. 

\newpage
\bibliographystyle{plain}
\bibliography{main}
\end{document}